\documentclass[english,twoside,a4paper,10pt]{article}

\usepackage[latin1]{inputenc}
\usepackage[T1]{fontenc}
\usepackage{amsmath}
\usepackage{amsfonts}
\usepackage{graphicx}
\usepackage{subfigure}
\usepackage{a4wide}
\usepackage{amssymb}
\usepackage{fancyhdr}
\usepackage{mathrsfs}
\usepackage[toc,page]{appendix}

\begin{document}

\title{\bf $k$-essence for warm inflation on the brane}
\author{ Lorenzo Sebastiani\footnote{E-mail address: l.sebastiani@science.unitn.it}
,\,\,\,
Ratbay Myrzakulov\footnote{Email address: rmyrzakulov@gmail.com}
\\
\\
\begin{small}
Department of General \& Theoretical Physics and Eurasian Center for
\end{small}\\
\begin{small} 
Theoretical Physics, Eurasian National University, Astana 010008, Kazakhstan
\end{small}\\
}

\date{}

\maketitle


\begin{abstract}
In this paper, $k$-inflation is analyzed in warm braneworld scenario. A general class of $k$-essence models with power-law kinetic term is investigated and weak and strong dissipation regimes are studied. Scalar perturbations and spectral index are derived. The results are discussed and applied to specific examples.
\end{abstract}



\section{Introduction}

Currently, it is well accepted the idea according to which the universe underwent a period of strong accelerated expansion after the Big Bang. The existence of the primordial acceleration was first proposed by 
 Guth and Sato~\cite{Guth, Sato} in 1981 and today several approaches to the inflationary paradigm exist (see Refs.~\cite{mukbook, Linde, revinflazione} for some reviews). Among them, it has been suggested that inflation may be the manifestation of some extra dimensions in which our universe is confined.
The idea (inspired by string theories) is that we live in a brane-universe embedded in an higher dimensional space-time, whose extra-dimensions (the bulk) are inaccessible to ordinary matter~\cite{brane1, brane2, brane3, brane4, brane5, brane6, brane7} but contribute to the gravitational field equations on the brane. In this respect, one of the most interesting case is given by five-dimensional bulk~\cite{brane8, brane9}, where
the fact that in the five dimensional space-time the fundamental Planck scale can be considerably
smaller than the effective Planck scale has 
profound consequences in
the field equations of our four dimensional universe~\cite{brane10, Maeda2, Maart, Langlois, Lopez}, leading to a significative modification of Friedmann equations.

In standard chaotic inflation, where accelerated expansion is driven by a canonical field dubbed ``inflaton'', some reheating mechanism for the particle production is necessary at the end of inflation.  
An alternative to this scenario is represented by warm inflation~\cite{warm1,warm2, warm3, warm4, warm5, warm6}, where the production of radiation and ultrarelativistic matter takes place during the early-time acceleration, since
the inflaton is coupled with radiation and the radiation energy density is not shifted away, leading to radiation dominated universe without invoking any reheating.

Recently, chaotic warm inflation has been analyzed in the braneworld scenario~\cite{Campo} and the aim of our work is to extend such a study to general classes of $k$-essence models, motivated by the fact that $k$-inflation, where
a scalar field with non-standard higher order kinetic terms supports the early-time acceleration~\cite{kess1, kess2, kess3}, is strictly connected with string theories and therefore braneworld cosmology~\cite{kbrane1, kbrane2, kbrane3} (for a review, see Ref.~\cite{kessrev} and references therein). 

The paper is organized in the following way. In Section {\bf 2}, a brief review about braneworld inflation and the mechanism of warm scenario is presented and $k$-essence Lagrangian is introduced. In Section {\bf 3}, a general class of $k$-essence models with power-law kinetic term is investigated in warm braneworld scenario for early-time acceleration. The strong and weak dissipation regimes are carefully analyzed with explicit examples. Section {\bf 4} is devoted to the study of scalar perturbations and to the derivation of spectral index. Final remarks are given in Section {\bf 5}.

We use units of $k_{\mathrm{B}} = c = \hbar = 1$ and denote the Planck Mass as
$M_{Pl} =1.2 \times 10^{19}$.

\section{Braneworld warm inflation}

If Einstein's gravity comes from an higher dimensional theory, the Einstein's equation in four dimension aquires new contributes at high energies. In a five dimensional hyperspace, the Einstein's equation can be written as
\begin{equation}
R_{\mu\nu}^{(5)}-\frac{g_{\mu\nu}^{(5)}}{2}R^{(5)}+\Lambda^{(5)}g_{\mu\nu}^{(5)}=\frac{8\pi}{M_5^2} T_{\mu\nu}^{(5)}\,,
\end{equation}
where the Ricci tensor $R_{\mu\nu}^{(5)}$ and the Ricci scalar $R^{(5)}$ are evaluated respect to the five-dimensional metric $g_{\mu\nu}^{(5)}$, $\Lambda^{(5)}$ is a cosmological constant and $T_{\mu\nu}^{(5)}$ is the stress-energy tensor of matter.  
The Plank scale $M_5$ in five dimension is related to the Plank Mass $M_{Pl}$ as
\begin{equation}
M_{Pl}=\sqrt{\frac{3}{4\pi}}\left(\frac{M_5^2}{\sqrt{\lambda}}\right)M_5\,,\label{p5}
\end{equation} 
where $0<\lambda$ is the tension of the 3-brane where the matter fields are confined.

The four-dimensional Einstein's field equations induced on the brane assumes the form~\cite{Maeda2}
\begin{equation}
G_{\mu\nu}=-\Lambda_\text{eff}g_{\mu\nu}+\frac{8\pi}{M_{Pl}^2} T_{\mu\nu}+\left(\frac{8\pi}{M_5^3}\right)^2
\pi_{\mu\nu}-E_{\mu\nu}\,,\label{Eineq}
\end{equation}
where $G_{\mu\nu}$ is the usual four dimensional Einstein's tensor,
\begin{equation}
G_{\mu\nu}=R_{\mu\nu}-\frac{g_{\mu\nu}}{2}R\,,
\end{equation}
$R_{\mu\nu}$ and $R$ being the Ricci tensor and the Ricci scalar in four dimension and $g_{\mu\nu}$ being the four dimensional metric tensor.
Moreover, $T_{\mu\nu}$ is the stress energy tensor of the brane and
$\pi_{\mu\nu}$ is a tensor quadratic in $T_{\mu\nu}$~\cite{Langlois},
\begin{equation}
\pi_{\mu\nu}=-\frac{1}{4}T_{\mu\alpha}T_{\nu}^{\alpha}+\frac{1}{12}T T_{\mu\nu}+\frac{1}{24}(3T_{\alpha\beta}T^{\alpha\beta}-T^2)g_{\mu\nu}\,,\label{pimunu}
\end{equation}
with $T=g^{\mu\nu}T_{\mu\nu}$ the trace of $T_{\mu\nu}$.
Finally, $E_{\mu\nu}$ 
encodes the effects of the five dimensional hyperspace graviton freedom degrees on the brane dynamics and
it is a projection of the five-dimensional Weyl tensor, while the effective cosmological constant in four dimension $\Lambda_\text{eff}$ is related to the cosmological constant $\Lambda^{(5)}$ in five dimension as
\begin{equation}
\Lambda_\text{eff}=\frac{4\pi}{M_5^3}\left(\frac{\Lambda^{(5)}M_5^3}{8\pi}+\frac{4\pi}{3M_5^3}\lambda^2\right)\,.
\end{equation}
For the sake of simplicity, following other proposals
and avoiding the discussion about the fine-tuning problem, we set
$\Lambda^{(5)}
=-2(4\pi)^2\lambda^2/(3M_5^6)$ such that 
\begin{equation}
\Lambda_\text{eff}=0\,.\label{lambdacond}
\end{equation}
Let us see now how the five dimensional gravitation effects emerge in flat Friedmann-Robertson-Walker (FRW) universe described by the metric
\begin{equation}
ds^2=-dt^2+a(t)^2d{\bf x}^2\,,\label{metric}
\end{equation}
with $a(t)\equiv a$  the scale factor depending on the cosmological time.

The first Friedmann equation is derived from (0,0)-component of (\ref{Eineq}) as~\cite{brane8, brane9},
\begin{equation}
H^2=\frac{8\pi}{3 M_{Pl}^2}\rho_{\text{eff}}\left(1+\frac{\rho_{\text{eff}}}{2\lambda}\right)+\frac{\mathcal E}{a^4}\,,
\label{F}
\end{equation}
where $H=\dot a/ a$ is the usual Hubble parameter. Here,
$\rho_{\text{eff}}$ is
the effective energy density of the universe contents and $\mathcal E$ an integration constant related to $E_{\mu\nu}$ which behaves as a radiation term and during the strong accelerated expansion of inflation is quickly shifted away, such that we may assume
\begin{equation}
\mathcal {E}=0\,.\label{EE}
\end{equation}
Thus, the brane effects induce a 
term proportional to $\rho_{\text{eff}}^2/(\lambda M_{Pl}^2)$ which modifies the classical Friedmann equation for the velocity.
This correction is related to the Plank scale in five dimension (\ref{p5}) as, 
\begin{equation}
\frac{\rho_\text{eff}}{2\lambda}=\frac{2\pi}{3}\left(\frac{\rho M_{Pl}^2}{M_5^6}\right)\,.
\end{equation}
Such a term becomes important at the scale of inflation, when $\lambda\ll\rho_{\text{eff}}$, while when $\rho_\text{eff}\ll\lambda$
we recover the results of General Relativity. To neglect the braneworld correction at the time of nucleosynthesis, it must be
$10\text{TeV}<M_5$, namely $(1 \text{MeV})^4<\lambda$~\cite{brane4}.\\
\\
Let us introduce the warm inflationary scenario. Generally speaking, due to the strong accelerated expansion of inflation, any radiation or matter field which is eventually present in the primordial universe is shifted away. As a consequence, a reheating process for the particle production is necessary at the end of inflation to recover the Friedmann universe. In the warm scenario, radiation and ultrarelativistic matter 
do not disappear during the early-time acceleration.
This is possible if there is a thermal contact between the radiation/ultrarelativistic matter fields
and 
the scalar field (inflaton) which drives the early-time acceleration~\cite{warm1, warm2, warm3}. Thus, the interaction between this components 
makes radiation not disappearing during the primordial expansion and at the end of inflation the reheating can be avoided. 

In our work, we will consider inflation driven by a general $k$-essence fluid whose Lagrangian reads~\cite{kess1, kess2},
\begin{equation}
\mathcal L_{(\phi)}=p(X, \phi)\,,\quad X=-\frac{g^{\mu\nu}\partial_\mu \phi\partial_\nu\phi}{2}\,,\label{kessence}
\end{equation}
where $p(X, \phi)$ is a function of the field $\phi$ and its kinetic energy $X$. The stress energy tensor associated to this Lagrangian reads
\begin{equation}
T^{\mu}_{(\phi)\nu}=(\rho(\phi, X)+p(\phi, X))u^{\mu}u_{\nu}+p(\phi, X)\delta^{\mu}_\nu\,,\quad u_\nu=\frac{\partial_\nu\phi}{\sqrt{2X}}\,,\label{Tfluid}
\end{equation}
where we identify $p(\phi, X)$ with the effective pressure of $k$-essence and $\rho(\phi, X)$ with its energy density. In the above expression, also the velocity field $u_\mu$ appears, and in homogeneous and isotropic universe one simply has $u_\mu=(1,0,0,0)$. 

The Lagrangian derivation leads to
\begin{equation}
\rho(\phi, X)=2X\frac{\partial p(\phi, X)}{\partial X}-p(\phi, X)\,,\label{rhophi}
\end{equation}
and one may also introduce an Equation of State (EoS) parameter for $k$-essence,
\begin{equation}
\omega(\phi, X)=\frac{p(\phi, X)}{\rho(\phi, X)}\,.\label{omegarhophi}
\end{equation}
This description is quite general. A fluid with energy density $\rho$ and isotropic pressure $p$ can be parametrized in such a way. For example, for the Equation of State $p=\rho^2/\rho_0$, one has $\rho(X)= (X^{1/4}-\rho_0)$ and $p(X)=(X^{1/4}-\rho_0)^2/\rho_0$; a radiation fluid with $p_{rad}=\rho_{rad}/3$ corresponds to the case $p(X)=X^2$. During inflation, we need fluids whose EoS parameter is close to minus one and in the corresponding field description we have the dependence on the field of the energy density and pressure. 
The simplest case is given by canonical scalar field where
\begin{equation}
\rho(\phi, X)=X+V(\phi)\,,\quad p=X-V(\phi)\,,\label{canonical}
\end{equation}
$V(\phi)$ being a potential of the scalar field. 

By taking into account the contribute of radiation also, the effective energy density present in the early universe is given by
\begin{equation}
\rho_{\text{eff}}=\rho(\phi, X)+\rho_{rad}\,,
\end{equation}
where $\rho_{rad}$ is the energy density of radiation. In warm inflation, a friction term relates the continuity equations of  $k$-essence and radiation. On flat FRW metric (\ref{metric}) we have
\begin{equation}
X=\frac{\dot\phi^2}{2}\,.
\end{equation}
Thus, we will work with the following conservation law  (the dot denotes the time derivative),
\begin{eqnarray}
\dot\rho(\phi, X)+3H\rho(\phi, X)(1+\omega(\phi, X))&=&-\mathcal Y f(\phi, X)\dot\phi^2\,,\nonumber\\
\dot\rho_{rad}+4H\rho_{rad}&=&\mathcal Y f(\phi, X)\dot\phi^2\,,\quad \quad 0<\mathcal Y\,, f(\phi, X)\,.
\label{conslaw}
\end{eqnarray}
The function $f(\phi, X)$ 
describes the decay of the $k$-essence field due to the interaction with radiation and it is
positive defined assuming the dimension of a mass, such that $\mathcal Y$ is dimensionless and is also positive to have $0<\rho_{rad}$ during inflation (when $\dot\rho_{rad}\simeq 0$).
The form of $\mathcal Y f(\phi, X)\dot\phi^2$ above was introduced in a phenomenological way in Ref.~\cite{warm1}  with canonical scalar field (\ref{canonical}) and $f(\phi, X)\equiv f(\phi)$ depending on the field only. Here, we generalized the function $f(\phi, X)$ by introducing the dependence on the kinetic term: in this way, the coupling can be adapted to be a friction term for $k$-essence field.\\
\\
During inflation, the effective energy density of the universe coincides with the one of $k$-essence, and 
from (\ref{F})--(\ref{EE}) we get
\begin{equation}
H^2\simeq\frac{8\pi}{3 M_{Pl}^2}\rho(\phi, X)\left(1+\frac{\rho(\phi, X)}{2\lambda}\right)\,.
\label{EOM1}
\end{equation}
Thus, from the first equation in (\ref{conslaw}) we obtain
\begin{equation}
\dot\phi\simeq-\frac{\partial_\phi\rho(\phi, X)}{3H\partial_X p(\rho, X)(1+r)}\,,
\quad r=\frac{\mathcal Y f(\phi, X)}{3 H\partial_X p(\phi, X)}\,.\label{Xeq}
\end{equation}
In this derivation, we assume that $|\ddot\phi\partial_X\rho(\phi,X)|\ll |3H\dot\phi \partial_Xp(\rho, X)|$: since from (\ref{rhophi}) we have $\partial_X\rho(\phi, X)=\partial_X p(\phi, X)+\dot\phi^2\partial_{XX}p(\phi, X)$, our assumption togheter with the usual slow-roll approximation $|\ddot\phi|\ll H|\dot\phi|$ is valid as long as $\dot\phi^2\partial_{XX}p(\phi, X)$ is on the same order or smaller than $\partial_X p(\phi, X)$. We also have introduced the useful adimensional parameter $r$. The strong/weak dissipation regimes correspond to $1\ll r$ and  $r\ll  1$, respectively.

Let us define the slow-roll parameter 
\begin{equation}
\epsilon:=-\frac{\dot H}{H^2}\simeq \left(\frac{M_{Pl}^2}{16\pi}\right)\frac{4\lambda(\lambda+\rho(\phi, X))}{(2\lambda+\rho(\phi, X))^2}\frac{1}{\rho^2(\phi, X)}\frac{(\partial_\phi\rho(\phi, X))^2}{\partial_X p(\phi, X)\left(1+r\right)}\,.\label{epsilon}
\end{equation}
The universe accelerates when $0<\epsilon<1$. For canonical scalar field (\ref{canonical}) we recover the result of Ref.~\cite{Campo}, namely $\epsilon<1$ when $p<\rho(\lambda+2\rho)/(3(\lambda+\rho))$. During inflation it must be $\epsilon\ll 1$, while acceleration ends when $\epsilon=1$.

Inflation is described by a (quasi) de Sitter solution with Hubble parameter almost a constant, such that $|\dot\phi/H|\ll 1$ and the field slowly changes. As a consequence, the energy density of radiation is almost a constant,
\begin{equation}
\rho_{rad}\simeq \frac{\mathcal Y f(\phi, X)\dot\phi^2}{4H}\,,
\label{23}
\end{equation}
and the thermal bath in primordial universe has the temperature $T=(\rho_{rad})^{1/4}$. Moreover,  the ratio between the energy density of radiation and the energy density of $k$-essence reads
\begin{equation}
\frac{\rho_{rad}}{\rho(\phi, X)}\simeq\frac{(2\lambda+\rho(\phi, X))}{4(\lambda+\rho(\phi, X))}\frac{\epsilon r}{\partial_X p(\phi, X)(1+r)}\,.\label{ratio}
\end{equation}
Thus, in the slow-roll approximation with $\epsilon\ll 1$, the universe results to be dominated by k-essence, but at the end of inflation, when $\epsilon\sim 1$, radiation emerges in the cosmological scenario without any reheating. 

To measure the amount of inflation, we need the $e$-folds number,
\begin{equation}
N:=\log\left[\frac{a(t_\text{f})}{a(t_\text{i})}\right]=\int^{t_\text{f}}_{t_\text{i}}H(t)dt\simeq
-\frac{8\pi}{M_{Pl}^2}\int^{\phi_\text{f}}_{\phi_\text{i}}\frac{\rho(\phi, X)}{2\lambda}(2\lambda+\rho(\phi, X))
\left(\frac{\partial_X p(\phi, X)(1+r)}{\partial_\phi\rho(\phi, X)}\right)d\phi\,.\label{N}
\end{equation}
Here, $t_\text{i,f}$ are the initial and final time of early-time acceleration and $\phi_{\text{i, f}}$ the related values of the field. Inflation is viable and brings to the thermalization of observable universe if $55<N<65$.

We point out that in the limit $\rho(\phi, X)\ll\lambda$, we recover all the formalism of classical warm scenario.
Let us analyze a general class of $k$-essence models in braneworld warm scenario, where inflation is realized in the limit $\lambda\ll\rho(\phi, X)$.

\section{Inflation in $k$-essence models with power-law kinetic term}

In this section, we will consider a class of $k$-essence models whose Lagrangian assumes the form
\begin{equation}
p(\phi, X)=F(X)-V(\phi)\,,
\end{equation}
where $F(X)$ and $V(\phi)$ are functions depending on $X$ and $\phi$ separately. $k$-essence models of this type are suitable to reproduce inflation: generally speaking, the potential $V(\phi)$ supports the de Sitter expansion as long as $\phi$ is almost a constant, while the term $F(X)$ makes ending inflation when $\dot\phi^2$ increases.

We can take,
\begin{equation}
F(X)=\beta X^\alpha\,,\quad 0<\beta\,, \quad 1\leq\alpha\,,
\end{equation}
where $\alpha$ is an adimensional parameter larger than one and $\beta$ is a positive parameter with dimension $[\beta]=[M_{Pl}]^{4(1-\alpha)}$.  In the limits $\alpha=\beta=1$ we recover the case of canonical scalar field (\ref{canonical}). From (\ref{rhophi}) we obtain
\begin{equation}
p(\phi, X)=\beta X^\alpha-V(\phi)\,,\quad \rho(\phi, X)=\beta(2\alpha-1) X^\alpha+V(\phi)\,.
\label{28}
\end{equation}
Inflation is supported by the potential $V(\phi)$ in the limit $\phi\rightarrow-\infty$, and we must require
\begin{equation}
X^\alpha \ll V(\phi\rightarrow-\infty)\,,\quad V(\phi\rightarrow 0^-)\ll X^\alpha\,.
\end{equation} 
Thus, the EoS parameter (\ref{omegarhophi}) behaves as
\begin{equation}
\omega(\phi\rightarrow-\infty, X\ll 1)\simeq -1\,,\quad
\omega(\phi\rightarrow 0^-, 1\ll X)\simeq \frac{1}{2\alpha-1}\,,
\end{equation}
from which one may infer a corresponding fluid description. In warm scenario, a friction term appears in (\ref{conslaw}). If we assume
\begin{equation}
f(\phi, X)= g(\phi)\alpha\beta X^{\alpha-1}\,,\label{gg}
\end{equation}
$g(\phi)$ being a positive dimensional function ($[g]=[M_{Pl}]$) of $\phi$ only, we get
\begin{equation}
\left(3H+\mathcal Y g(\phi)\right)\alpha\beta\, 2^{1-\alpha}\dot\phi^{2\alpha}\simeq-V'(\phi)\dot\phi\,,
\label{consex1}
\end{equation}
where the prime denotes the derivative respect to the field and we used the slow-roll approximation
$|\ddot\phi|\ll H |\dot\phi|$. The strong/weak dissipation regimes correspond to  $3H\ll \mathcal Y g(\phi)$ and $\mathcal Y g(\phi)\ll 3H$, respectively, or, in terms of
\begin{equation}
r=\frac{\mathcal Y g(\phi)}{3H}\,,
\end{equation}
to $1\ll r$ and $r\ll 1$.
During inflation, the field mainly contributes to the effective energy density of the universe. Thus, from (\ref{EOM1}) with $\lambda\ll \rho(\phi, X)$, we have
\begin{equation}
H^2\simeq\frac{8\pi}{3 M_{Pl}^2}\frac{V(\phi)^2}{2\lambda}\,.
\end{equation}
Moreover, the field slowly moves according to (\ref{Xeq}), which is a consequence of (\ref{consex1}) and leads to
\begin{equation}
\dot\phi\simeq \left(\frac{-2^{\alpha-1}V'(\phi)}{3H(1+r)\beta\alpha}\right)^{\frac{1}{2\alpha-1}}\,.
\end{equation}
Thus, the slow-roll parameter (\ref{epsilon}) is evaluated as
\begin{equation}
\epsilon\simeq\left(\frac{24\pi}{M_{Pl}^2\lambda}\right)^{\frac{1-\alpha}{1-2\alpha}}\frac{
M_{Pl}^2\lambda V'(\phi)^2
\left( 
-\frac{V'(\phi)}{\alpha\beta V(\phi)(1+r)}
\right)^{2(\alpha-1)/(1-2\alpha)}
}{4\pi\alpha\beta(1+r)V(\phi)^3}\,.\label{epsilonex}
\end{equation}
For canonical scalar field $\epsilon=M_{Pl}^2\lambda V'(\phi)^2/(4\pi(1+r)V(\phi)^3)$ according with Ref.~\cite{Campo}. From  (\ref{ratio}) we obtain
\begin{equation}
\frac{\rho_{rad}}{\rho(\phi, X)}\simeq\frac{\epsilon r}{4\partial_X p(\phi, X)(1+r)}\propto r\lambda^{1/(2\alpha-1)}\frac{\left(- M_{Pl}V'(\phi)/\beta\right)^{2/(2\alpha-1)}}{V(\phi)^{(2\alpha+1)/(2\alpha-1)}(1+r)^{2/(2\alpha-1)}}\,,
\end{equation}
which is vanishing as long as $r\left(-V'(\phi) M_{Pl}/(1+r)\right)^{2/(2\alpha-1)}\ll V(\phi)^{(2\alpha+1)/(2\alpha-1)}$, such that $\rho_{rad}\ll\rho(\phi, X)$ during inflation. At the end of inflation, the $\epsilon$ slow-roll parameter is on the order of the unit and the model enters (or remains) in the strong dissipation regime with $1\ll r$. Thus, the energy density of radiation becomes dominant and we recover the Friedmann universe.

We will investigate now different cases of dissipation regime during accelerated phase.

\subsection{Strong dissipation regime}

In analogy with the original proposal of Ref.~\cite{warm1}, we may assume a dependence on $\phi$ of the coupling term between $k$-essence and radiation, which is encoded in the function $g(\phi)$ introduced in (\ref{gg}). This choice permits to obtain a strong dissipation regime during and after inflation under the requirement
$1\ll r$, namely
\begin{equation}
3H\ll\mathcal Y g(\phi)\,.\label{sd}
\end{equation}
The $\epsilon$ slow-roll parameter (\ref{epsilonex}) reads
\begin{equation}
\epsilon\simeq
2^{(1-\alpha)/(1-2\alpha)} M_{Pl}^{(4-4\alpha)}
\sqrt{\frac{3 M_{Pl}^2\lambda}{\pi}}
\frac{V'(\phi)^2}{2\mathcal Y\alpha\beta g(\phi)V(\phi)^2}
\left(-\frac{V'(\phi)}{\mathcal Y\alpha\beta g(\phi) M_{Pl}^{(4\alpha-2)}}\right)^{2(\alpha-1)/(1-2\alpha)}\,.
\label{epsilons}
\end{equation}
Thus, in the limit $\alpha\rightarrow 0$ and $\beta\rightarrow 0$, one has for canonical scalar field
\begin{equation}
\epsilon\simeq\sqrt{\frac{3M_{Pl}^2\lambda}{4\pi}}\frac{V'(\phi)^2}{\mathcal Y g(\phi) V(\phi)^2}\,,\quad
\alpha=\beta=1\,,
\end{equation}
in agreement with Ref.~\cite{Campo}. On the other hand, in the limit $\alpha\rightarrow+\infty$, we obtain
\begin{equation}
\epsilon\simeq-M_{Pl}^2\sqrt{\frac{3 M_{Pl}^2\lambda}{2\pi}}\frac{V'(\phi)}{V(\phi)^2}\,,\quad 1\ll\alpha\,,
\label{e2}
\end{equation}
and the explicit dependence on the coupling disappears in higher order $k$-essence models. The ratios between the energy density of inflation and $k$-essence result to be
\begin{equation}
\frac{\rho_{rad}}{\rho(\phi, X)}\simeq\sqrt{\frac{3M_{Pl}^2\lambda}{\pi}}\frac{V'(\phi)^2}{8\mathcal Y g(\phi)V(\phi)^2}\,,\quad \alpha=\beta=1\,,\label{ratio1}
\end{equation}
and
\begin{equation}
\frac{\rho_{rad}}{\rho(\phi, X)}\simeq M_{Pl}^4\sqrt{\frac{3M_{Pl}^2\lambda}{\pi}}\frac{\mathcal Y g(\phi)}{4V(\phi)^2}\,,\quad 1\ll\alpha\,.\label{ratio2}
\end{equation}
Finally,
the $e$-folds numbers (\ref{N}) are derived as
\begin{equation}
N\simeq-\mathcal Y\sqrt{\frac{4\pi}{3M_{Pl}^2\lambda}}\int^{\phi_\text{f}}_{\phi_\text{i}}
\frac{g(\phi)V(\phi)}{V'(\phi)}\,,\quad \alpha=\beta=1\,,\label{N1}
\end{equation}
and
\begin{equation}
N\simeq\frac{1}{M_{Pl}^2}\sqrt{\frac{2\pi}{3\lambda M_{Pl}^2}}\int^{\phi_\text{f}}_{\phi_\text{i}}V(\phi)\,,
\quad 1\ll\alpha\,.\label{N2}
\end{equation}
As an example,
let us take the power-law forms for the potential and the dissipation term, namely
\begin{equation}
V(\phi)=\frac{\gamma(-\phi)^n}{n}
\,,\quad
g(\phi)=g_0 (-\phi)^m\,,\quad 1\leq n\,, m\,,
\end{equation}
where
$\gamma$ and
$g_0$ are positive dimensional parameters and $n\,, m$ positive numbers. Inflation is realized for $\phi\rightarrow-\infty$, 
\begin{equation}
H^2\simeq\frac{8\pi}{3 M_{Pl}^2}\frac{\gamma^2(-\phi)^{2n}}{2 n^2 \lambda}\,.\label{H1}
\end{equation}
Thus, the strong dissipation regime (\ref{sd}) takes place if $n< m$. By using (\ref{epsilons}) with (\ref{N}) in the limit $|\phi_\text{f}|\ll|\phi_{\text{i}}|$ we obtain
\begin{equation}
\epsilon\simeq\frac{n}{N}\frac{(2\alpha-1)}{(m+2\alpha(n+1)-2n)}\,,
\end{equation}
where the $e$-folds number is given by
\begin{equation}
N=\sqrt{\frac{\pi}{3M_{Pl}^2\lambda}}\frac{(2\alpha-1)}{n(m+2\alpha(n+1)-2n)}\frac{(-\phi_\text{i})
^\frac{m-2n+2\alpha(n+1)}{(2\alpha-1)}
}{A_0}\,,\quad
A_0=\frac{2^{\frac{1-\alpha}{1-2\alpha}}}{2g_0\mathcal Y\alpha\beta}\left(\frac{\gamma}{g_0\mathcal Y\alpha\beta}\right)^{\frac{2(1-\alpha)}{2\alpha-1}}\,.
\end{equation}
We see that if $1\leq\alpha$ and $n<m$, we get $1\ll N$. On the other hand, in the limit $\alpha=\beta=1$ we recover the result of Ref.~\cite{Campo},
\begin{equation}
\epsilon\simeq\frac{n^2}{2g_0\mathcal Y}\sqrt{\frac{3M_{Pl}^2\lambda}{\pi}}\frac{1}{\phi_\text{i}^{m+2}}\,,\quad\alpha=\beta=1\,.
\end{equation}
In our example, the ratio between the energy density of radiation and $k$-essence reads
\begin{equation}
\frac{\rho_{rad}}{\rho(\phi, X)}\simeq
g_0\mathcal Y n^2\sqrt{\frac{3M_{Pl}^2\lambda}{\pi}}\frac{A_0^2}{2^{\frac{1}{1-2\alpha}}}\frac{1}{(-\phi)
^{\frac{2+4n(\alpha-1)-m(3\alpha-3)}{2\alpha-1}}}\,,\label{sdratio}
\end{equation}
and in the given limits we recover (\ref{ratio1})--(\ref{ratio2}). Therefore, the exponent of ($-\phi$) must be positive to have $\rho_{rad}\ll\rho(\phi, X)$ for large (negative) values of the field. For example, in the limit $\alpha=1$, every positive value of $m$ is viable, while in the limit $1\ll\alpha$ we must have $n<m<2n$ to realize warm inflation in strong dissipation regime. We can easily verify that at the end of inflation, when $\phi\rightarrow 0^-$, the energy density of radiation becomes dominant respect to the one of $k$-essence.

\subsection{Weak dissipation regime}

Here, we will analyze the weak dissipation regime during the accelerated phase,
\begin{equation}
\mathcal Y g(\phi)\ll 3H\,.
\end{equation}
In this case, only when the early-time acceleration ends we enter in the strong dissipation regime and the energy of $k$-essence is converted into radiation. The $\epsilon$ slow-roll parameter (\ref{epsilonex}) is given by
\begin{equation}
\epsilon\simeq
(24\pi)^{\frac{1-\alpha}{1-2\alpha}} M_{Pl}^{(4-4\alpha)}
\frac{M_{Pl}^2\lambda V'(\phi)^2}{4\pi\alpha\beta V(\phi)^3}
\left(-\frac{V'(\phi)\sqrt{\lambda M_{Pl}^2}}{\alpha\beta V(\phi) M_{Pl}^{(4\alpha-2)}}\right)^{2(\alpha-1)/(1-2\alpha)}\,.
\label{epsilonw}
\end{equation}
In the limits $\alpha=1$ and $\beta=1$, one has
\begin{equation}
\epsilon\simeq
\frac{M_{Pl}^2\lambda V'(\phi)^2}{4\pi V(\phi)^3}
\,,
\end{equation}
\begin{equation}
\frac{\rho_{rad}}{\rho(\phi, X)}\simeq
\frac{\mathcal Y g(\phi) (\lambda M_{Pl}^2)^{3/2}V'(\phi)^2}{32\sqrt{3}\pi^{3/2} V(\phi)^4}\,,
\end{equation}
\begin{equation}
N\simeq-\frac{4\pi}{M_{Pl}^2\lambda}
\int^{\phi_\text{f}}_{\phi_\text{i}}\frac{V(\phi)^2}{V'(\phi)}\,,\quad \alpha=\beta=1\,.
\end{equation}
Therefore, at the end of inflation, in strong dissipation regime, the ratio between the energy density of inflation and $k$-essence
is given by (\ref{ratio1}) again. On the other hand, it is observed that, in the limit $1\ll\alpha$, we obtain the same asymptotic behaviors (\ref{e2}, \ref{ratio2}, \ref{N2}) of strong dissipation case for higher order $k$-essence models.

Let us consider as an example,
\begin{equation}
V(\phi)=\frac{\gamma(-\phi)^n}{n}
\,,\quad
g(\phi)=g_0\,,\quad 1\leq n\,,\label{57}
\end{equation}
where $\gamma$ and $g_0$ are dimensional constants and $n$ a positive number. The Hubble parameter is given by (\ref{H1}) and the weak dissipation regime is realized for large and negative values of the field.

By taking the limit
$|\phi_\text{f}|\ll|\phi_{\text{i}}|$ and by
plugging  (\ref{N}) in  (\ref{epsilonw}) , we derive
\begin{equation}
\epsilon\simeq\frac{n}{N}\frac{(2\alpha-1)}{(2\alpha(n+1)-n)}\,,
\end{equation}
where the $e$-folds is
\begin{equation}
N=
\frac{1}{\lambda M_{Pl}^2}\frac{B_0}{n^2}\frac{2\alpha-1}{\left(2\alpha(n+1)-n\right)}(-\phi_\text{i})^{\frac{2\alpha+n(2\alpha-1)}{(2\alpha-1)}}
\,,\quad
B_0=4\pi\alpha\beta\gamma\left(24\pi\right)^{\frac{\alpha-1}{1-2\alpha}}
\left(
\frac{n\sqrt{\lambda M_{Pl}^2}}{\alpha\beta}
\right)
^{\frac{2(\alpha-1)}{2\alpha-1}}
\,.
\end{equation}
In the limit $1\ll\alpha$ we have $\epsilon\simeq n/\left(N(n+1)\right)$ like in the strong dissipation case analyzed in the preceding subsection: this is a consequence of the fact that the $\epsilon$ parameter, the $e$-folds number and therefore their match acquire the same forms in the given limit.

Now the ratio of the energy density of radiation and $k$-essence is derived as
\begin{equation}
\frac{\rho_{rad}}{\rho(\phi, X)}\simeq
\frac{g_0 n^4\mathcal Y\sqrt{\pi}(\lambda M_{Pl}^2)^{3/2}}{2\sqrt{3}B_0^2}\frac{1}{(-\phi)^{\frac{n(4\alpha-2)+2}{(2\alpha-1)}}}\,,
\end{equation}
which is small for large values of the field, such that expansion is driven by $k$-essence. At the end of inflation, when the Hubble parameter decreases and $3H<\mathcal Y g_0$, we enter in the strong dissipation phase and we recover (\ref{sdratio}) with $m=0$: thus, for small values of the field, the energy density of radiation becomes dominant and one finds the usual Friedmann cosmology.

\section{Scalar perturbations}

Let us consider the scalar perturbations in FRW metric (\ref{metric}) by using the Newton's gauge,
\begin{equation}
ds^2=-(1+2\Phi(t, {\bf x}))dt^2+a^2(t)(1-2\Psi(t, {\bf x}))\delta_{ij}dx^i dx^j\,,\quad i,j=1,2,3\,.
\end{equation} 
Here, $|\Phi(t, {\bf x})\,, \Psi(t, {\bf x})|\ll 1$ are functions of the cosmological time and the spatial coordinates such that $g^{00}\simeq-1+2\Phi(t, {\bf x})$ and $g^{ii}\simeq (1+2\Psi(t, {\bf x}))/a(t)^2$.
The perturbations on the $k$-essence will bring to
\begin{equation}
\delta\rho(\phi, X)\simeq \partial_\phi\rho(\phi, X)\delta\phi(t, {\bf x})+\partial_X\rho(\phi, X)\delta X(t, {\bf x})\,,\quad 
\delta X(t, {\bf x})\simeq \dot\phi^2\left(-\Phi(t, {\bf x})+\frac{\delta\dot\phi(t, {\bf x})}{\dot\phi}\right)\,,
\label{deltarho} 
\end{equation}
while the perturbed velocity field introduced in (\ref{Tfluid}) reads
\begin{equation}
u(t, {\bf x})=\left(1+\Phi(t, {\bf x}), {\bf \partial} \frac{\delta\phi(t, {\bf x})}{\dot\phi}\right)\,.
\end{equation}
Since from the $(i,j)$ components of (\ref{Eineq}), when $i, j=1,2,3\,,i\neq j$, at the first order in  $\delta\phi(\phi, X)$ we obtain $[\Phi(t, {\bf x})-\Psi(t, {\bf x})]\simeq 0$, in the following we will pose
\begin{equation}
\Psi(t, {\bf x})=\Phi(t, {\bf x})\,.
\end{equation}
The $(0,0)$-component (on shell) and the $(0,i)$-components of (\ref{Eineq}) are derived as
\begin{equation}
2\left[\frac{\triangle\Phi(t, {\bf x})}{a(t)^2}-3H\left(\dot\Phi(t, {\bf x})+H\Phi(t, {\bf x})\right)\right]=\frac{8\pi}{M_{Pl}^2}\delta\rho(\phi, X)\left(1+\frac{\rho(\phi, X)}{\lambda}\right)\,,
\label{65}
\end{equation}
\begin{equation}
2\partial_i
\left(\dot\Phi(t, {\bf x})+H\Phi(t, {\bf x})\right)=
\partial_i
\left[
\frac{8\pi}{M_{Pl}^2}\left(1+\frac{\rho(\phi, X)}{\lambda}\right)\left[\left(\rho(\phi, X)+p(\phi, X)\right)\frac{\delta\phi(t, {\bf x})}{\dot\phi}
-\psi_r(t, {\bf x})
\right]
\right]\,,\label{66}
\end{equation}
where we have taken into account relations (\ref{p5}, \ref{lambdacond}, \ref{EE}) and we used (\ref{pimunu}) at the first order, namely
\begin{equation}
\pi_{oi}=\frac{\rho}{6}(\rho(\phi, X)+p(\phi, X))\left(\frac{\partial_i\delta\phi(t, {\bf x})}{\dot\phi}\right)\,.
\end{equation}
Moreover, in Eq.~(\ref{65}) we considered the energy density of $k$-essence dominant respect to the one of radiation, while in Eq.~(\ref{66}) $\psi_r(t, {\bf x})$ represents the energy fluss of radiation which obeys to the law~\cite{covcina},
\begin{equation}
\dot\psi_r(t, {\bf x})+3H\psi_r(t, {\bf x})=-\frac{4}{3}\rho_{rad}\Phi(t, {\bf x})
-\frac{\delta\rho_{rad}(t, {\bf x})}{3}-\mathcal Y f(\phi, X)\dot\phi\delta\phi(t, {\bf x})\,.
\end{equation}
In order to perturb the continuity equation of $k$-essence in (\ref{conslaw}), we must introduce the covariant formalism~\cite{Bardeen},
\begin{equation}
\nabla_\mu T^{\mu \nu}_{(\phi)}=-\mathcal Y f(\phi, X)\left(\partial_\alpha\phi u^\alpha\right)
\partial^\nu\phi\,,\label{last}
\end{equation}
to get
(on the background equation),
\begin{eqnarray}
&&\delta\dot\rho(\phi, X)+3H(\delta\rho(\phi, X)+\delta p(\phi, X))-\frac{(\rho(\phi, X)+p(\phi,X))\triangle\delta\phi(t, {\bf x})}{a^2\dot\phi}
\nonumber\\&&
-3\dot\Phi(t, {\bf x})(\rho(\phi, X)+p(\phi,X))=\nonumber\\&&
\mathcal Y f(\phi, X)\dot\phi\left[3\dot\phi\Phi(t, {\bf x})-2\delta\dot\phi(t, {\bf x})\right]-\mathcal Y\dot\phi^2\left[
\frac{\partial f(\phi, X)}{\partial X}\delta X(t, {\bf x})+
\frac{\partial f(\phi, X)}{\partial \phi}\delta \phi(t, {\bf x})
\right]\,.
\end{eqnarray}
In a similar way, from the continuity equation of radiation one obtains 
\begin{eqnarray}
&&\delta\dot\rho_{rad}(t, {\bf x})+4H\delta\rho_{rad}(t, {\bf x})
+\frac{\triangle\psi_r(t,{\bf x})}{a^2}
-4\dot\Phi(t, {\bf x})\rho_{rad}=\nonumber\\&&
-\mathcal Y f(\phi, X)\dot\phi\left[3\dot\phi\Phi(t, {\bf x})-2\delta\dot\phi(t, {\bf x})\right]+\mathcal Y\dot\phi^2\left[
\frac{\partial f(\phi, X)}{\partial X}\delta X(t, {\bf x})+
\frac{\partial f(\phi, X)}{\partial \phi}\delta \phi(t, {\bf x})
\right]\,,
\end{eqnarray}
and for canonical scalar field (\ref{canonical}) we recover the perturbed equations of Ref.~\cite{Campo}.\\
\\
From Eq.~(\ref{deltarho}), by using the conservation law in (\ref{conslaw}) we get
\begin{equation}
\delta\rho(\phi, X)=\dot\phi^2\rho_X \left[-\Phi(t, {\bf x})+\frac{d}{dt}\left(\frac{\delta\phi(t, {\bf x})}{\dot\phi}\right)\right]-\frac{3H(\rho(\phi, X)+p(\phi, X))\delta\phi(t,{\bf x})}{\dot \phi}
-\mathcal Y f(\phi, X)\dot\phi\delta\phi(t, {\bf x})\,.
\label{deltarho2}
\end{equation}
Since from Eq.~(\ref{66}) with (\ref{conslaw})--(\ref{EOM1}) we obtain the following relation
\begin{eqnarray}
\frac{d}{dt}\left[
H\frac{\delta\phi(t, {\bf x})}{\dot\phi}+\Phi(t, {\bf x})
\right]&=&
\left[
H\frac{d}{dt}\left(\frac{\delta\phi(t, {\bf x})}{\dot\phi}\right)-H\Phi(t, {\bf x})
\right]\nonumber\\&&
-\frac{4\pi}{3H M_{Pl}^2}\left(\mathcal Y f(\rho, X)\dot\phi\delta\phi(t, {\bf x})+3H\psi_r(t, {\bf x})\right)\left(1+\frac{\rho}{\lambda}\right)\,,
\end{eqnarray}
from Eqs.~(\ref{65})--(\ref{66}) together with (\ref{deltarho2}) we can derive
\begin{eqnarray}
\frac{\triangle\Phi(t, {\bf x})}{a^2}&=&
\frac{4\pi}{M^2_{Pl}}\left[
\frac{\dot\phi^2\rho_X(\phi, X)}{H}
\frac{d}{dt}\left(H\frac{\delta\phi(t, {\bf x})}{\dot\phi}+\Phi(t, {\bf x})\right)
\right.\nonumber\\&&\left.
\hspace{-3cm}+(\mathcal Y f(\rho, X)\dot\phi\delta\phi(t, {\bf x})+3H\psi_r(t, {\bf x}))\left(
\frac{(4\pi)\dot\phi^2\rho_X(\phi, X)}{3H^2 M_{Pl}^2}\left(1+\frac{\rho(\phi, X)}{\lambda}\right)
-1
\right)
\right]\left[1+\frac{\rho(\phi, X)}{\lambda}\right]\,,\label{74}
\end{eqnarray}
\begin{eqnarray}
\frac{d}{dt}\left(\frac{a\Phi(t, {\bf x})}{H}\right)&=&
\frac{4\pi a}{M_{Pl}^2}
\frac{(\rho(\phi, X)+p(\phi, X))}{H^2}\left(H\frac{\delta\phi(t, {\bf x})}{\dot\phi}+\Phi(t, {\bf x})\right)
\left(1+\frac{\rho(\phi, X)}{\lambda}\right)\nonumber\\&&
+\frac{4\pi a}{3H^2 M_{Pl}^2}\left(\frac{\mathcal Y f(\phi, X)\dot\phi^2\Phi(t, {\bf x})}{H}-3H\psi_r(t, {\bf x})\right)\left(1+\frac{\rho(\phi, X)}{\lambda}\right)\,.\label{75}
\end{eqnarray}
By plugging (\ref{75}) in (\ref{74}), and by using the slow roll approximations $|\dot H/H^2|$, $|\ddot H/(H\dot H)|$, $|\dot\phi/(H \phi) |$, $|\ddot\phi/(H\dot\phi)|\ll 1$, in the limit $\rho_{rad}\ll \rho(\phi, X)$ one has the closed equation
\begin{eqnarray}
\frac{\triangle\Phi(t, {\bf x})}{a^2}&\simeq&\frac{1}{c_s^2}(\ddot\Phi(t, {\bf x})+H\dot\Phi(t, {\bf x})+\dot H\Phi(t, {\bf x}))-(\dot\Phi(t, {\bf x})+H\Phi(t, {\bf x}))\left(\frac{\frac{d}{dt}\left(\rho(\phi,X)+p(\phi, X)\right)}{c_s^2(\rho(\phi, X)+p(\phi, X))}\right)\nonumber\\&&
+
\frac{H\Phi(t, {\bf x})}{c_s^2}\left(
\frac{\dot H}{H}-\frac{\dot\rho(\phi, X)}{(\lambda+\rho(\phi, X))}
\right)
+H\Phi(t, {\bf x})\left(\frac{\mathcal Y f\dot\phi^2}{\rho(\phi, X)}\right)\left(\frac{\lambda+\rho(\phi, X)}{2\lambda+\rho(\phi, X)}\right)
\,,\label{76}
\end{eqnarray}
where we have assumed $\dot\rho_{rad}\simeq 0$ and $\rho_{rad}\simeq \mathcal Y f(\phi, X)\dot\phi^2$ and we have considered $\delta\rho_{rad}(t, {\bf x})$, $\dot\psi_r(t, {\bf x})\simeq 0$. 
Here, we introduced the sound speed
\begin{equation}
c_s^2=\frac{(\rho(\phi, X)+p(\phi, X))}{\dot\phi^2\rho_X(\phi, X)}=\frac{p_X(\phi, X)}{\rho_X}\,.
\end{equation}
For canonical scalar field $c_s=1$, but for $k$-essence $c_s<1$: for instance, in the case of (\ref{28}), we get 
$c_s=1/(2\alpha-1)$.
By redefining
\begin{equation}
\Phi(t, {\bf x})=\frac{4\pi}{M_{Pl}^2}\frac{(\rho(\phi, X)+p(\phi, X))^{1/2}}{\sqrt{a}}u(t, {\bf x})\,,
\end{equation}
equation~(\ref{76}) reads
\begin{eqnarray}
&&
\frac{\triangle u(t, {\bf x})}{a^2}\simeq\frac{\ddot u(t, {\bf x})}{c_s^2}+\frac{u(t, {\bf x})}{c_s^2}
\left(
\frac{3\dot H}{2}-\frac{H^2}{4}+\frac{\frac{d^2}{dt^2}(\rho(\phi, X)+p(\phi, X))}{2(\rho(\phi, X)+p(\phi, X))}
+\frac{\frac{d}{dt}(\rho(\phi, X)+p(\phi, X))}{2(\rho(\phi, X)+p(\phi, X))}H
\right.
\nonumber\\&&
\left.
\hspace{-1cm}
-\frac{3}{4}\left(\frac{\frac{d}{dt}(\rho(\phi, X)+p(\phi, X))}{\rho(\phi, X)+p(\phi, X)}\right)^2
-\frac{H\dot\rho(\phi, X)}{(\lambda+\rho(\phi, X)}
\right)
+H u(t, {\bf x})\left(
\frac{\mathcal Y f(\phi, X)\dot\phi^2}{\rho(\phi, X)}
\right)\left(\frac{\lambda+\rho(\phi, X)}{2\lambda+\rho(\phi, X)}\right)\,.
\end{eqnarray}
By decomposing the perturbation in Fourier modes $u_k(t, {\bf x})=u_k(t)\exp\left[i {\bf k} {\bf x}\right]$, we have that for short-wave perturbations with $Ha \ll |k|$,
\begin{eqnarray}
u_k(t)&\simeq& c_k\left[
\sqrt{\frac{a}{c_s}}
\text{e}^{i\int\frac{H a}{2k c_s}\frac{\mathcal Y f(\phi, X)\dot\phi^2 dt}{\rho(\phi, X)}\left(\frac{\lambda+\rho(\phi, X)}{2\lambda+\rho(\phi, X)}\right)}
\right]
\exp\left[
\pm ik\int \frac{c_s dt}{a}
\right]\,,
\nonumber\\
\Phi_k(t)&\simeq& \frac{4\pi}{M_{Pl}^2} c_k\frac{(\rho(\phi, X)+p(\phi, X))^{1/2}}{\sqrt{c_s}}
\left[
\text{e}^{+i\int\frac{H a}{2k c_s}\frac{\mathcal Y f \dot\phi^2 dt}{\rho}\left(\frac{\lambda+\rho}{2\lambda+\rho}\right)}
\right]
\exp\left[
\pm ik\int \frac{c_s dt}{a}
\right]\,,\label{80}
\end{eqnarray} 
$c_k$ being an integration constant. On the other hand, for long-wave perturbations with $|k|\ll H a$, one obtains
\begin{eqnarray}
u_k(t)&\simeq& \frac{A_k M_{Pl}^2}{4\pi} \Gamma(t)
\frac{\sqrt{a}}{\sqrt{\rho(\phi, X)+p(\phi, X)}}\left[\frac{\rho(\phi, X)+\lambda}{\lambda}\right]\left[1-\frac{H}{a}\int a dt\right]\,,\nonumber\\
\Phi_k(t)&\simeq& A_k\Gamma(t) \left[\frac{\rho(\phi, X)+\lambda}{\lambda}\right]\left[1-\frac{H}{a}\int a dt\right]\,,\label{81}
\end{eqnarray}
where $A_k$ is an integration constant. In this expressions, $\Gamma(t)$ encodes the contribution from the coupling term between $k$-essence and radiation and can be derived as
\begin{equation}
\Gamma(t)\simeq
\exp\left[
-\int dt \frac{ \mathcal Y f(\phi, X)\dot\phi^2 }{\rho(\phi, X)}
\left(\frac{\lambda+\rho(\phi, X)}{2\lambda+\rho(\phi, X)}\right)\right]\,.
\end{equation}
We also have in slow roll approximation,
\begin{equation}
\frac{1}{a}\int a dt=\frac{1}{a}\int\frac{da}{H}=\frac{1}{H}-\frac{1}{a}\int\frac{da}{H}\frac{d}{dt}\left(\frac{1}{H}\right)dt\simeq\frac{1}{H}\left(1+\frac{\dot H}{H^2}\right)\,.\label{83}
\end{equation}
The amplitude of a scalar perturbation on the sound horizon crossing ($c_s k\simeq H a$) depends on the initial condition on $c_k$ in (\ref{80}). From the quantum field theory of perturbations one must pose~\cite{mukbook},
\begin{equation}
c_k=-\frac{i M^{3/2}}{k^{3/2}}\,,
\end{equation}
Thus, on the sound horizon, in (\ref{81}) we set
\begin{equation}
A_k=-\frac{1}{k^{3/2}\sqrt{c_s}}\left(\frac{4\pi i}{\sqrt{M_{Pl}}}\right)\frac{\sqrt{\rho(\phi,X)+p(\phi, X)}}{\Gamma(t)}
\left(\frac{\lambda}{\lambda+\rho(\phi, X)}\right)\left(-\frac{H^2}{\dot H}\right)|_{c_s k\simeq H a}\,,
\end{equation}
where we used (\ref{83}). As a consequence, at the end of inflation, the variance of the power spectrum of long-wave perturbations reads
\begin{eqnarray}
\delta_\Phi^2(k)&\equiv& \frac{4(\rho(\phi, X)+p(\phi, X))|u_k|^2 k^3}{a M_{Pl}^7}\nonumber\\
&=&
 \frac{4(\rho(\phi, X)+p(\phi, X))}{c_s M_{Pl}^4 \Gamma^2(t)}\left(\frac{\lambda}{\rho(\phi, X)+\lambda}\right)^2\left(-\frac{H^2}{\dot H}\right)^2|_{c_s k\simeq H a}
\left[1-\frac{H}{a}\int a dt\right]^2\,,
\end{eqnarray}
where we have taken the limits of Friedmann universe with $\Gamma(t)=1$ and $\lambda\rightarrow\infty$.
In the specific, during radiation era with $a(t)\sim t^{1/2}$, the supercurvature perturbations are frozen such that
\begin{equation}
\delta_\Phi^2(k)=
 \frac{16(\rho(\phi, X)+p(\phi, X))}{9 c_s M_{Pl}^4 \Gamma^2(t)}\left(\frac{\lambda}{\rho(\phi, X)+\lambda}\right)^2\left(-\frac{H^2}{\dot H}\right)^2|_{c_s k\simeq H a}
\end{equation}
Then, the spectral index $n_s$ reads
\begin{eqnarray}
(n_s-1)&=&\frac{d \ln \delta_\Phi^2(k)}{d \ln k}\simeq
\frac{d \ln \delta_\Phi^2(k)}{d \ln a_k}
\nonumber\\&=&
\frac{4\dot H}{H^2}-\frac{2\Ddot H}{H\dot H}+\frac{\frac{d}{dt}(\rho(\phi, X)+p(\phi, X))}{H(\rho(\phi, X)+p(\phi, X))}
-\frac{\dot\rho(\phi, X)}{H(\lambda+\rho(\phi, X))}
-\frac{\dot c_s}{H c_s}\nonumber\\
&&+\frac{8\rho_{rad}}{\rho(\phi, X)}\left(\frac{\lambda+\rho(\phi, X)}{2\lambda+\rho(\phi, X)}\right)
\,,
\end{eqnarray}
where we taken into account (\ref{23}).
If we remove the contribute of the dissipation ($\mathcal Y=0$), 
by using (\ref{conslaw})--(\ref{EOM1}) we have 
\begin{equation}
n_s-1=\frac{4\dot H}{H^2}-\frac{\ddot H}{H \dot H}-\frac{\dot c_s}{H c_s}\,,
\end{equation}
and we can verify that when the sound speed is independent on the time the spectral index is given by the usual formula $(n_s-1)=-6\epsilon+2\eta$ with $\epsilon=-\dot H/H^2$ and $\eta=\epsilon-\ddot H/(2H\dot H)$ valid for chaotic inflation from canonical scalar field~\cite{Maart}.

As an example, we can consider the $k$-essence model in (\ref{57}) in the limit $1\ll\alpha$ and $n=1$. Since inflation is realized in weak dissipation regime and
$(\rho(\phi, X)+p(\phi, X))\simeq-\dot\rho(\phi, X)/3H$, by taking into account that
$\ddot H/(H\dot H)=(d\epsilon/dN)/(2\epsilon)-\epsilon$,
the spectral index can be easily calculated as
\begin{equation}
(n_s-1)\simeq -\frac{2}{N}\left(1-\frac{1}{\sqrt{2}}\left(\frac{\mathcal Y g_0 M_{Pl}^2}{\gamma}\right)\right).
\end{equation}
In this case, in order to satisfy the last Planck satellite results~\cite{Planck, WMAP} for $N\simeq 60$, it must be $\mathcal Y g_0 M_{Pl}^2/(\sqrt{2}\gamma)\ll 1$.

We conclude by noting that, since the ratio of tensor to scalar power spectrum amplitudes $\delta_h^2(k)/\delta_\Phi^2(k)$ is proportional to the sound speed $c_s$, for $k$-essence with $c_s< 1$
it is strongly suppressed, easily satisfying  the Planck constraints (it is the case of (\ref{57}) with $1\ll\alpha$, $c_s\rightarrow 0$) .

\section{Conclusions}

In this paper, we have analyzed $k$-inflation in warm braneworld cosmology. In this theory, the mechanism of braneworld inflation (in our specific five-dimensional case, Randall-Sundrum models) is combined with warm scenario. 
Braneworld cosmology is a string-inspired theory and in such a context $k$-essence is one of the possible description of the early-time acceleration. The Lagrangian of $k$-essence includes higher order kinetic terms and may have an effective  
fluid representation, such that it appears as a quite general theory. Moreover, the introduction of a friction term in the continuity equation of $k$-essence and the consequent coupling with radiation permits to maintain constant the energy density of radiation/ultrarelativistic matter during primordial acceleration, avoiding any reheating at the beginning of Friedmann expansion. We observe that, due to the higher order kinetic term of $k$-essence, the coupling with radiation must explicitly depend on such kinetic term in order to play the role of a friction. 

We furnished the formalism to study inflation and we investigated a class of $k$-essence models, whose formulation is quite general. In our $k$-Lagrangian, a power-law kinetic term appears, keeping separate the dependence on the field to support the (quasi) de Sitter expansion with a suitable potential. Strong and weak dissipation regimes are analyzed and discussed. 

In the last part of the paper, we studied the scalar perturbations in Friedmann-Robertson-Walker space-time deriving the spectral index of the theory. The corresponding formula leads to the standard results in the limits of General Relativity and we can see that the coupling between radiation and $k$-essence introduces a contribute proportional to the ratio between their energy densities.

Note that FRW dynamics may be changed if one considers brane in different bulk (for instance, in $AdS-BH$ bulk ~\cite{AdSBHbulk}).
An other interesting application of the topic is the inclusion 
of modified gravity with $k$-essence (see the approach used in Ref.~\cite{MGk}).
Other works about generalizations of braneworld cosmology can be found in Refs.~\cite{superbrane1, superbrane2}.

\end{document}